\documentclass[twocolumn,showpacs,superscriptaddress,amsmath,amssymb,aps,pra]{revtex4-1}
\usepackage{graphicx}
\usepackage{CJKutf8}
\usepackage{dcolumn}
\usepackage{amsmath,bm}
\usepackage{epstopdf}
\usepackage[mathlines]{lineno}
\usepackage{color}
\usepackage[colorlinks=true,linkcolor=blue,citecolor=magenta,urlcolor=cyan]{hyperref}

\begin{document}
\title{Charge noise induced spin dephasing in a nanowire double quantum dot with spin-orbit coupling}
\author{Rui\! Li~(\begin{CJK}{UTF8}{gbsn}李睿\end{CJK})}
\email{ruili@ysu.edu.cn}
\affiliation{Key Laboratory for Microstructural Material Physics of Hebei Province, School of Science, Yanshan University, Qinhuangdao 066004, China}

\begin{abstract}
Unexpected fluctuating charge field near a semiconductor quantum dot has severely limited the coherence time of the localized spin qubit. It is the interplay between the spin-orbit coupling and the asymmetrical confining potential in a quantum dot, that mediates the longitudinal interaction between the spin qubit and the fluctuating charge field. Here, we study the $1/f$ charge noise induced spin dephasing in a nanowire double quantum dot via exactly solving its eigen-energies and eigenfunctions. Our calculations demonstrate that the spin dephasing has a nonmonotonic dependence on the asymmetry of the double quantum dot confining potential. With the increase of the potential asymmetry, the dephasing rate first becomes stronger very sharply before reaching to a maximum, after that it becomes weaker softly. Also, we find that the applied external magnetic field contributes to the spin dephasing, the dephasing rate is strongest at the anti-crossing point $B_{0}$ in the double quantum dot.
\end{abstract}
\date{\today}
\maketitle

\section{Introduction}
The coherence of a quantum bit (qubit), an interesting phenomenon originating from the superposition of quantum states in quantum mechanics, has many applications in quantum computing and quantum information processing~\cite{nielsen2002quantum,ladd2010quantum}. Electron spin, localized in semiconductor quantum dot~\cite{loss1998quantum,hanson2007spins}, is an excellent qubit candidate due to its convenience for both manipulation~\cite{petta2005coherent,koppens2006driven} and scalability~\cite{burkard1999coupled,shulman2012demonstration} in experiments. Both a fast spin manipulation and a long spin coherence times are required to achieve a reliable quantum computer~\cite{buluta2011natural,ladd2010quantum}. Spin dephasing induced by unexpected environment noise is the primary obstacle limiting the potential applications of the spin qubit~\cite{astafiev2004quantum,you2007low,bermeister2014charge,kha2015do}.   

$1/f$ charge noise~\cite{paladino2014noise}, an interesting environment noise whose spectrum density has a $1/f$ distribution, has been observed in various quantum nano-systems~\cite{jung2004background,bylander2011noise,kuhlmann2013charge,chan2018assessment}. Recently, spin dephasing induced by $1/f$ charge noise was observed in a Si quantum dot integrated with micromagnet~\cite{kawakami2016gate,yoneda2018}. The longitudinal slanting field created by the nearby micromagnet mediates a longitudinal spin-charge interaction that gives rise to the spin pure dephasing~\cite{Li_2019}. A detailed theoretical investigation on the underlying physics of the spin dephasing not only helps clarify the mysterious $1/f$ spectrum distribution of the noise, but also can guide us how to improve the spin coherence time.

The spin-orbit coupling (SOC)~\cite{bychkov1984oscillatory}, intrinsically presented in III-V semiconductor quantum dot, can mediate a transverse spin-charge interaction which leads to a well-known effect called electric-dipole spin resonance~\cite{golovach2006electric,Nowack1430,lirui2013controlling,PhysRevB.99.014308}. Recently, a longitudinal spin-charge interaction is demonstrated in a single quantum dot with both SOC and asymmetrical confining potential~\cite{lirui2018a}. The longitudinal spin-charge interaction gives rise to the spin pure dephasing due to the $1/f$ charge noise~\cite{lirui2018a}. Since a double quantum dot (DQD) is usually used to produce a large transverse spin-charge interaction near the anti-crossing point, with applications in both spin manipulation~\cite{kawakami2014electrical} and cavity quantum electrodynamics~\cite{hu2012strong,Petersson:2012aa}, it is of practical importance to study the $1/f$ charge noise induced spin dephasing in this system.

In this paper, the spin dephasing in a spin-orbit coupled nanowire DQD modeled by an infinite double square well is explored in detail. We find that a little asymmetry in the DQD confining potential, e.g., several fractions of a milli-electron-volt (meV) in the potential difference or several nanometers (nm) in the width difference between the left and the right dots of the DQD, can give rise to a remarkable spin dephasing. This would be instructive and meaningful to the quantum computing architecture based on semiconductor quantum dot, because it is almost impossible to produce an exactly symmetrical quantum dot confining potential in experiments. A nonmonotonic dependence of the spin dephasing on the potential asymmetry is demonstrated. We also find the applied external magnetic field contributes to the longitudinal spin-charge interaction, hence there is a sharp dip at the anti-crossing point $B_{0}$ when the spin coherence time $T^{*}_{2}$ is plotted as a function of the magnetic field.

The paper is organized as follows. In Sec.~\ref{sec_II}, we give the DQD model we are interested in. In Sec.~\ref{sec_III}, we calculate exactly the energy spectrum and the corresponding eigenfunctions of the DQD. In Sec.~\ref{sec_IV}, the spin dephasing as a function of the potential asymmetry is studied in detail. In Sec.~\ref{sec_V}, we demonstrate the spin dephasing also has a magnetic field dependence. At last, we give a summary in Sec.~\ref{sec_VI}.

\section{\label{sec_II}The nanowire DQD}
\begin{figure}
\includegraphics{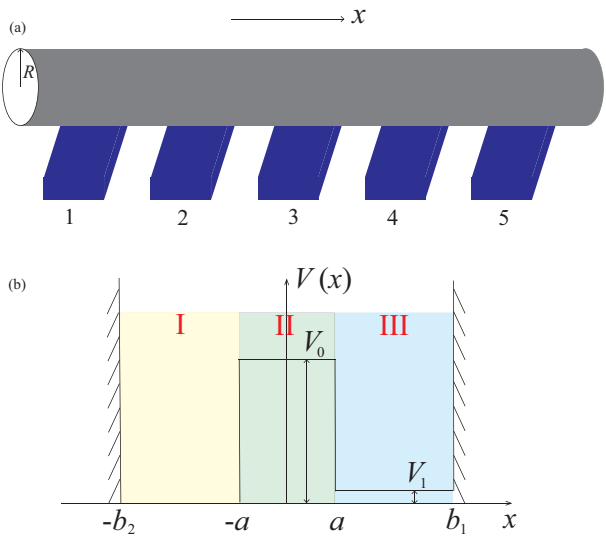}
\caption{\label{Fig_model}(a) Schematic realization of a DQD by placing a InSb nanowire of radius $R$ on a set of metallic gates. The five gates below the nanowire are used to produce a double-well confining potential along the longitudinal direction, i.e., $x$ direction. (b) The asymmetrical infinite double square well used to mimic the longitudinal confining potential of the nanowire DQD. The potential asymmetry can be tuned by varying either the potential difference $V_{1}$ between the dots or the width $b_{1}$ of the right dot.}
\end{figure}

Semiconductor DQD with one electron confined has been investigated for a long time~\cite{PhysRevLett.91.226804,PhysRevB.74.155433,PhysRevB.72.155410,gorman2005charge,khomitsky2012spin}. In the early development of quantum computing, a quantum dot charge qubit was proposed using the two localized charge states in a DQD~\cite{PhysRevLett.74.4083}. However, charge qubit is sensitive to external unexpected charge noise, such that its coherence time is usually very short~\cite{PhysRevLett.91.226804,petersson2010quantum}. On the other hand, quantum dot spin qubit based on the electron's spin degree of freedom has broader applications, due to its less sensitivity to charge noise~\cite{witzel2006quantum,yao2006theory,koppens2008spin}. Besides, it is easy to achieve a strong electric-dipole spin resonance in a DQD, which is useful for the single spin qubit manipulation.

Here, we are interested in a quasi one-dimensional (1D) nanowire DQD. A simple physical realization of the nanowire DQD is given in Fig.~\ref{Fig_model}(a). Via controlling the electric potentials on the five gates below the nanowire, we achieve a double-well confining potential [see Fig.~\ref{Fig_model}(b)] along the longitudinal direction, i.e., the $x$ direction, of the nanowire. For a more detailed experimental realization of this process, see Refs.~\cite{nadj2010spin,nadj2012spectroscopy}. The transverse dimension of the DQD is given by the nanowire radius $R$. 
Note that in the limit of strong transverse confinement, i.e., the radius $R$ ($\sim10$ nm) of the nanowire is much less than the longitudinal dot size $b_{2}$ ($\sim60$ nm), the kinematics of the electron in the nanowire DQD is approximately quasi 1D~\cite{trif2008spin}. The transverse motions of the electron (perpendicular to the wire) are approximately frozen, and we only need to consider the longitudinal motion along the $x$ direction.

In order to show explicitly the underlying physics of the spin dephasing and its dependence on the DQD parameters, here the DQD confining potential  is modeled by an infinite double square well [see Fig.~\ref{Fig_model}(b)]. A conduction electron of the semiconductor material is localized in this double-well potential. The DQD Hamiltonian under investigation reads
\begin{equation}
H=\frac{p^{2}}{2m}+\alpha\sigma^{z}p+\Delta\sigma^{x}+V(x),\label{Eq_model}
\end{equation}
where $m$ is the effective electron mass, $\alpha$ is the Rashba SOC strength~\cite{bychkov1984oscillatory}, $\Delta=g\mu_{B}B/2$ is half of the Zeeman splitting in the presence of an external magnetic field ${\bf B}$ applied in the $x$ direction, and the general double-well potential with a little bit asymmetry reads [see Fig.~\ref{Fig_model}(b)]
\begin{equation}
V(x)=\left\{\begin{array}{ll}
0,&{\rm region~I}:~-b_{2}<x<-a,\\
V_{0},&{\rm region~II}:~-a<x<a,\\
V_{1},&{\rm region~III}:~a<x<b_{1},\\
\infty,&{\rm elsewhere}.
\end{array}
\right.
\end{equation}

As emphasized in our previous study~\cite{lirui2018a}, the interplay between the SOC and the asymmetrical quantum dot confining potential can mediate a longitudinal spin-charge interaction, which gives rise to the spin pure dephasing. Here the  potential asymmetry of the DQD can be tuned by varying either the parameter $V_{1}$ or the parameter $b_{1}$ [see Fig.~\ref{Fig_model}(b)]. 

Our first step is to find the eigen-energies and the corresponding eigenfunctions of our DQD model (\ref{Eq_model}). Note that the boundary condition is important for determining the eigen-energies of a quantum system~\cite{landau1965quantum}. For the double-well model we are considering, the boundary condition explicitly reads~\cite{lirui2018a}
\begin{eqnarray}
\Psi(-b_{2})&=&0,~\Psi(\pm\,a-0)=\Psi(\pm\,a+0),\nonumber\\
\Psi(b_{1})&=&0,~\Psi'(\pm\,a-0)=\Psi'(\pm\,a+0),\label{Eq_boundary}
\end{eqnarray}
where $\Psi(x)=[\Psi_{1}(x),\Psi_{2}(x)]^{\rm T}$ is the quantum dot eigenfunction to be determined and $\Psi'(x)$ is its first derivative. Here, $\Psi_{1,2}(x)$ are the two components of the eigenfunction due to the spin degree of freedom.

\begin{table}
	\centering
	\caption{\label{tab}The parameters of a InSb DQD used in our calculations}
	\begin{ruledtabular}
		\begin{tabular}{ccccc}
			$m/m_{0}$\footnote{$m_{0}$ is the free electron mass}&$\alpha$~(eV \AA)~\cite{winkler2003spin}&$g$&$B_{0}$~(T)&~\\
			$0.0136$&$1.05$&$50.6$&$0.05$&~\\
			$a$~(nm)&$b_{1}$~(nm)&$b_{2}$~(nm)&$V_{0}$~(meV)&$V_{1}$~(meV)\\
			$10$&$55\sim60$&$60$&$50$&$0\sim0.9$
		\end{tabular}
	\end{ruledtabular}
\end{table}

There is a strong Rashba SOC in the InSb material~\cite{winkler2003spin,nadj2010spin,nadj2012spectroscopy}, such that here we only study in detail the spin dephasing in a InSb nanowire DQD, the physics in other III-V materials, such as InAs and GaAs, would be similar. In our following calculations, unless otherwise specified, the DQD parameters are taken from Table~\ref{tab}.

\section{\label{sec_III}Energy spectrum and Eigenfunctions of the DQD}
\begin{figure}
\includegraphics{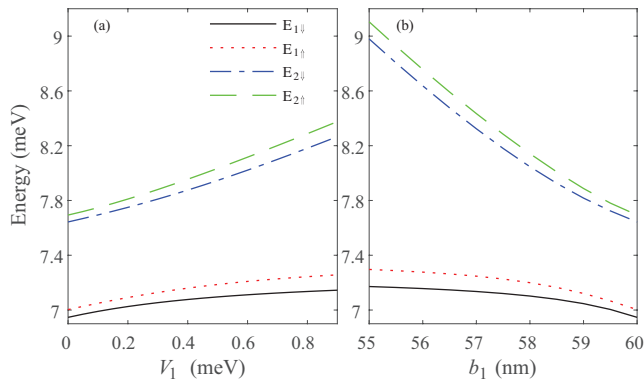}
\caption{\label{Fig_ES}The lowest four energy levels, labeled with indices $|1\!\Downarrow\rangle$, $|1\!\Uparrow\rangle$, $|2\!\Downarrow\rangle$, and $|2\!\Uparrow\rangle$, in the nanowire DQD. (a) The energy levels as a function of the potential difference $V_{1}$ between the left and right dots. Here $b_{1}=60$ nm. (b) The energy levels as a function of the right dot width $b_{1}$. Here $V_{1}=0$ meV.}
\end{figure}

Following the standard method for tackling the square well problem with SOC~\cite{lirui2018energy,lirui2018the,liu2018spin,lirui2018a,bulgakov2001spin,tsitsishvili2004rashba}, we have obtained the energy spectrum and the corresponding eigenfunctions of the DQD exactly (the detailed method is given in Appendix~\ref{appendix_a}). By introducing $12$ coefficients $c_{i}$ ($i=1,\ldots,12$) to be determined, we expand the eigenfunction in terms of the bulk wavefunctions [see Eq.~(\ref{Eq_eigenfunction})]~\cite{lirui2018a,lirui2018the}. Note that the eigenfunction is a piecewise function with respect to coordinate $x$, and we have three sub-regions for coordinate $x$ in the DQD. The boundary condition (\ref{Eq_boundary}) actually contains $12$ sub-equations. Therefore, substituting the eigenfunction in (\ref{Eq_boundary}) with that given in (\ref{Eq_eigenfunction}), we obtain a matrix equation ${\bf M}\cdot{\bf C}=0$, where ${\bf M}$ is a $12\times12$ matrix with $E$ being its variable (for details see Appendix~\ref{appendix_a}). The solution of the transcendental equation ${\rm det}(\bf M)$=0, an implicit equation of $E$, gives us the energy spectrum of the DQD. Once an eigen-energy, e.g., $E_{n}$, is obtained, we can solve the corresponding coefficients $c^{n}_{i}$ by combining the equation ${\bf M}\cdot{\bf C}=0$ with the normalization condition $\int\,dx\Psi^{\dagger}_{n}(x)\Psi_{n}(x)=1$. Substituting the solved $c^{n}_{i}$ into Eq.~(\ref{Eq_eigenfunction}), we then have the eigenfunction $\Psi_{n}(x)$ with eigenvalue $E_{n}$. 

Note that because spin is not a good quantum number in Hamiltonian (\ref{Eq_model}), strictly speaking, the lowest Zeeman sublevels in the DQD encode a spin-orbit qubit~\cite{nadj2010spin,nadj2012spectroscopy}, not a spin qubit. Therefore, in this paper, the spin qubit should be understood as a spin-orbit qubit. Obviously, this qubit contains a little orbital degree of freedom of the electron in the DQD in addition to the spin degree of freedom, such that the spin-orbit qubit can respond to both the external magnetic and electric fields~\cite{berg2013fast}. 
 
In a semiconductor DQD, the lowest four energy levels, labeled with $|1\!\Downarrow\rangle$, $|1\!\Uparrow\rangle$, $|2\!\Downarrow\rangle$, and $|2\!\Uparrow\rangle$, respectively, are relevant to the design of the spin-orbit qubit. The lowest four energy levels as a function of the potential asymmetry of the DQD are shown in Fig.~\ref{Fig_ES}. In Fig.~\ref{Fig_ES}(a), the asymmetry is tuned by varying the potential difference $V_{1}$ between the left and right dots. It should be noted that with the increase of $V_{1}$, the level splitting of the spin-orbit qubit $E_{1\!\Uparrow}-E_{1\!\Downarrow}$ becomes larger. While in Fig.~\ref{Fig_ES}(b), the asymmetry is tuned by changing the width $b_{1}$ of the right dot. Similarly, with the decrease of $b_{1}$, the level splitting of the qubit $E_{1\!\Uparrow}-E_{1\!\Downarrow}$ also becomes larger.

\begin{figure}
	\includegraphics{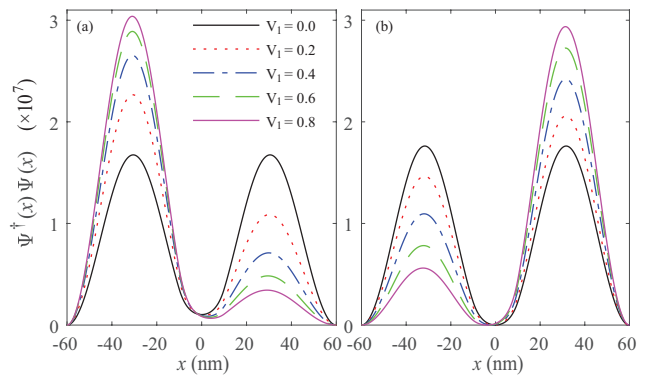}
	\caption{\label{Fig_WFV1}The probability density distribution of the first eigenstate $|1\!\Downarrow\rangle$ (a) and the third eigenstate $|2\!\Downarrow\rangle$ (b) in the DQD with different potential difference $V_{1}$ between the left and right dots. Here $b_{1}=60$ nm.}	
\end{figure} 

\begin{figure}
	\includegraphics{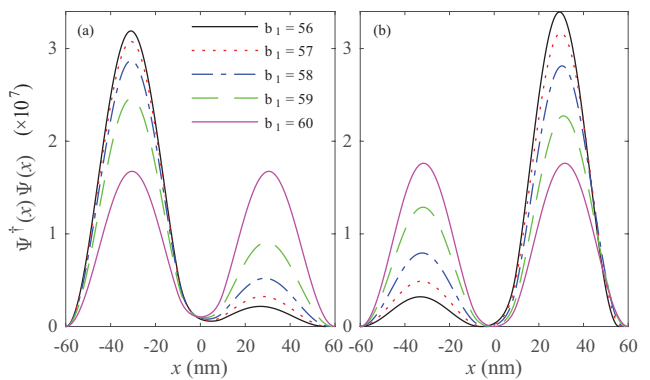}
	\caption{\label{Fig_WFb1}The probability density distribution of the first eigenstate $|1\!\Downarrow\rangle$ (a) and the third eigenstate $|2\!\Downarrow\rangle$ (b) in the DQD with different right dot width $b_{1}$. Here $V_{1}=0$ meV.}
\end{figure}

Once the energy spectrum of the DQD is obtained, we can calculate the corresponding eigenfunctions. The probability density distributions of states $|1\!\Downarrow\rangle$ and $|2\!\Downarrow\rangle$ with various potential differences $V_{1}$ and various right dot widths $b_{1}$ are shown in Figs.~\ref{Fig_WFV1} and~\ref{Fig_WFb1}, respectively. The probability density distribution of state $|1\!\Uparrow\rangle$ is similar to that of $|1\!\Downarrow\rangle$, and the probability density distribution of state $|2\!\Uparrow\rangle$ is similar to that of $|2\!\Downarrow\rangle$, are not shown here. If the DQD confining potential is symmetrical, i.e., $V_{1}=0$ meV in Fig.~\ref{Fig_WFV1} or $b_{1}=60$ nm in Fig.~\ref{Fig_WFb1}, the probability density of the eigenstate also has a symmetrical distribution with respect to the $x=0$ axis. Once the potential asymmetry is presented, the symmetrical probability density distribution is broken. Also, with the increase of the potential asymmetry in the DQD, i.e., via increasing $V_{1}$ or shortening $b_{1}$, the probability density of the ground state $|1\!\Downarrow\rangle$ becomes more localized to one of the dot [see Figs.~\ref{Fig_WFV1}(a) and~\ref{Fig_WFb1}(a)], i.e., the probability density in the left dot is larger than that in the right dot. This induces interesting effect related to the results shown in Fig.~\ref{Fig_ES}. We can understand as follows. First, when the probability density of the ground state becomes more localized to the left dot, the effective DQD size $x_{\rm size}$ becomes smaller too. One can imagine in the strong asymmetrical limit, the DQD would become a single quantum dot, there is almost no probability density distribution of the state in the right dot. Second, the spin-orbit effect in the quantum dot is roughly characterized by the ratio $x_{\rm size}/x_{\rm so}$, where $x_{\rm so}=\hbar/(m\alpha)$ is the spin-orbit length. Hence, with the increase of the potential asymmetry, the qubit level splitting $E_{1\!\Uparrow}-E_{1\!\Downarrow}$, roughly proportional to $g\mu_{B}B{\rm exp}\left[-(x_{\rm size}/x_{\rm so})^{2}\right]$~\cite{trif2008spin,lirui2013controlling,lirui2018the}, becomes larger as illustrated in Fig.~\ref{Fig_ES}.

\section{\label{sec_IV}Spin pure dephasing}
\begin{figure}
	\includegraphics{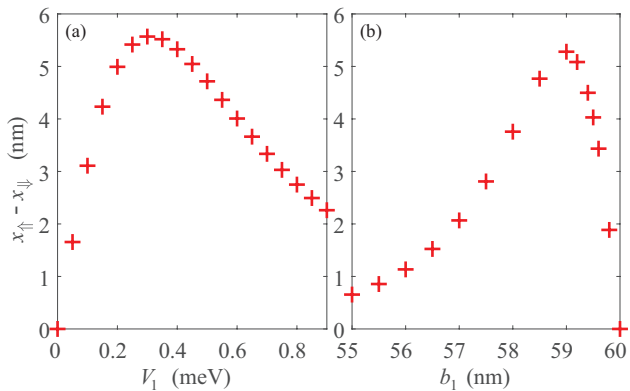}
	\caption{\label{Fig_Xoperator}The element of the longitudinal spin-charge interaction, represented by $x_{\Uparrow}-x_{\Downarrow}$, as a function of the DQD potential asymmetries. (a) The difference $x_{\Uparrow}-x_{\Downarrow}$ as a function of the potential difference $V_{1}$ between the left and right dots. Here $b_{1}=60$ nm. (b) The difference $x_{\Uparrow}-x_{\Downarrow}$ as a function of the right dot width $b_{1}$. Here $V_{1}=0$ meV.}
\end{figure}

\begin{figure}
	\includegraphics{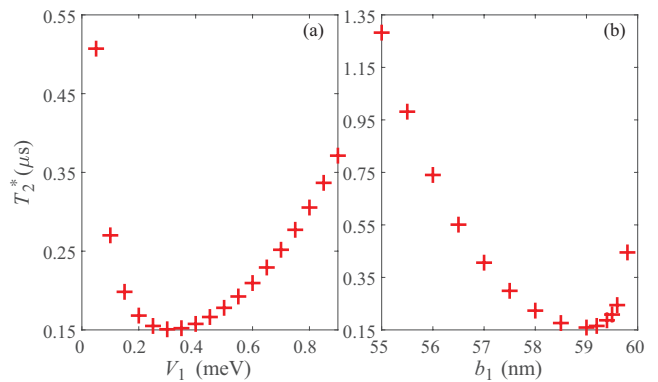}
	\caption{\label{Fig_dephasing}The qubit coherence time $T^{*}_{2}$ as a function of the DQD potential asymmetries. (a) The $T^{*}_{2}$ as a function of the potential difference $V_{1}$ between the left and right dots. Here $b_{1}=60$ nm. (b) The $T^{*}_{2}$ as a function of the right dot width $b_{1}$. Here $V_{1}=0$ meV.}
\end{figure}

The transverse interaction between a quantum dot spin qubit and an external driving electric field has been demonstrated a decade ago. The representative example is the quantum dot electric-dipole spin resonance~\cite{golovach2006electric,Nowack1430,lirui2013controlling,PhysRevB.99.014308,nowak2013spin,romhanyi2015subharmonic}. However, the transverse spin-charge interaction only leads to potential spin relaxation~\cite{huang2014electron}. Recently, a longitudinal spin-charge interaction, which induces spin pure dephasing, has been demonstrated in a single quantum dot with both SOC and asymmetrical confining potential~\cite{lirui2018a}.

$1/f$ charge noise, the spectrum density of which is inversely proportional to the noise frequency, has attracted considerable interests over many decades~\cite{dutta1981low,bermeister2014charge,kha2015do,weissman1988noise,paladino2014noise}. Here we study the $1/f$ charge noise induced spin pure dephasing in a nanowire DQD. Following the method in deriving the interaction Hamiltonian between a two-level system and a bosonic noise~\cite{scully1999quantum}, here we construct the model of the spin qubit interacting with the $1/f$ charge noise. The fluctuating charge field can be expressed as ${\bf E}=\sum_{k}\Xi_{k}\hat{e}_{k}(b_{k}+b^{\dagger}_{k})$,
where $\Xi_{k}$ is the charge field in the wavevector space, and $\hat{e}_{k}$ is a unit vector. Hence, the total Hamiltonian describing the qubit-noise interaction reads
\begin{equation}
H_{\rm tot}=H+ex\cos\Theta\sum_{k}\Xi_{k}(b_{k}+b^{\dagger}_{k})+\sum_{k}\hbar\omega_{k}b^{\dagger}_{k}b_{k},\label{eq_spin_noise}
\end{equation}
where $\cos\Theta=\hat{x}\cdot\hat{e}_{k}$, for simplicity, we have assumed $\hat{e}_{k}=\hat{e}$ does not depend on $k$. When we project $H_{\rm tot}$ into the Hilbert subspace spanned by the qubit basis states: $|1\!\Uparrow\rangle$ and $|1\!\Downarrow\rangle$, the longitudinal qubit-noise interaction is characterized by the difference between  $x_{\Uparrow}=\langle1\!\Uparrow|x|1\!\Uparrow\rangle$ and $x_{\Downarrow}=\langle1\!\Downarrow|x|1\!\Downarrow\rangle$~\cite{lirui2018a}. In this case, the total Hamiltonian can be reduced to
\begin{eqnarray}
H_{\rm tot}&=&\frac{E_{1\Uparrow}-E_{1\Downarrow}}{2}\tau^{z}+\sum_{k}\hbar\omega_{k}b^{\dagger}_{k}b_{k}+\nonumber\\
&&\sum_{k}e\Xi_{k}\left(\frac{x_{\Uparrow}-x_{\Downarrow}}{2}\tau^{z}+\frac{x_{\Uparrow}+x_{\Downarrow}}{2}\right)(b_{k}+b^{\dagger}_{k})\cos\Theta,\nonumber\\
\end{eqnarray}
where $\tau^{z}=|1\!\Uparrow\rangle\langle1\!\Uparrow|-|1\!\Downarrow\rangle\langle1\!\Downarrow|$ is the Pauli $z$ matrix.
As can be seen from the above equation, the emergence of the qubit phase noise, i.e., the qubit-noise interaction is longitudinal, is due to the following general reason. The average value of the electric-dipole operator $ex$ in one Zeeman sublevel $|1\!\Downarrow\rangle$ is different from that in the other Zeeman sublevel $|1\!\Uparrow\rangle$.

The difference $x_{\Uparrow}-x_{\Downarrow}$ is an important quantity for characterizing the longitudinal spin-charge interaction. In Fig.~\ref{Fig_Xoperator}, we show $x_{\Uparrow}-x_{\Downarrow}$ as a function of the potential asymmetries in the DQD. Interestingly, we find that $x_{\Uparrow}-x_{\Downarrow}$ has a nonmonotonic dependence on the potential asymmetries. Note that we can tune large the potential asymmetry by either increasing $V_{1}$ or shortening $b_{1}$ [see Fig.~\ref{Fig_model}(b)]. There is a critical potential difference $V^{c}_{1}$ [see Fig.~\ref{Fig_Xoperator}(a)] or a critical right dot width $b^{c}_{1}$ [see Fig.~\ref{Fig_Xoperator}(b)], at which $x_{\Uparrow}-x_{\Downarrow}$ becomes maximal. Also, with the increase of the potential asymmetry, the difference $x_{\Uparrow}-x_{\Downarrow}$ first becomes larger and larger sharply before reaching to its maximum, after that $x_{\Uparrow}-x_{\Downarrow}$ gets smaller and smaller softly. It should be noted that the calculated $x_{\Uparrow}-x_{\Downarrow}$ here is in the nanometer (nm) scale, much larger than that ($10^{-2}$ nm) mediated by a slanting magnetic field in a Si quantum dot~\cite{Li_2019}. It is also easier to produce a large $x_{\Uparrow}-x_{\Downarrow}$ in the DQD in comparison with the results in a single quantum dot~\cite{lirui2018a}. Last, at $V_{1}=0$ meV in Fig.~\ref{Fig_Xoperator}(a) or $b_{1}=60$ nm in Fig.~\ref{Fig_Xoperator}(b), the difference $x_{\Uparrow}-x_{\Downarrow}$ is exactly zero. This is because when we choose these parameters the model (\ref{Eq_model}) has a $Z_{2}$ symmetry as already discussed in Ref.~\cite{lirui2018a}.

Let us discuss on the underlying mechanism leading to the nonmonotonic behavior shown in Fig.~\ref{Fig_Xoperator}. The longitudinal spin-charge interaction, represented by $x_{\Uparrow}-x_{\Downarrow}$, is proportional to both the SOC strength and the degree of the asymmetry in the DQD~\cite{lirui2018a}. The nonmonotonic behavior is simply owing to the reason that, with the increase of the potential asymmetry, the relative SOC strength in the DQD decreases, such that there must exist a critical site ($V^{c}_{1}$ or $b^{c}_{1}$) where the combined effect of the SOC and the asymmetrical potential becomes maximal. Next, we explain why the relative SOC is inversely proportional to the asymmetry of the DQD. With the increase of the asymmetry, the effective DQD size $x_{\rm size}$ becomes smaller, the relative SOC strength, characterized by the ratio $x_{\rm size}/x_{\rm so}$~\cite{trif2008spin,lirui2013controlling,lirui2018the}, hence becomes smaller too.

The phase coherence of the spin qubit is described by the off-diagonal element of the qubit density matrix. If we model the phase coherence as $\left|\rho_{\Uparrow\Downarrow}(t)/\rho_{\Uparrow\Downarrow}(0)\right|={\rm exp}[-\langle\Gamma(t)\rangle_{\Theta}]$, the decay factor has the following exact expression~\cite{palma1996quantum}
\begin{equation}
\langle\Gamma(t)\rangle_{\Theta}=\frac{(x_{\Uparrow}-x_{\Downarrow})^{2}}{2(b_{2}-a)^{2}}\int^{\omega_{\rm max}}_{\omega_{\rm min}}S(\omega)\frac{\sin^{2}(\omega\,t/2)}{(\omega/2)^{2}},\label{Eq_decay}
\end{equation}
where 
\begin{equation}
S(\omega)=\sum_{k}\frac{e^{2}\Xi^{2}_{k}(b_{2}-a)^{2}k_{B}T}{\hbar^{2}\omega}\delta(\omega-\omega_{k})\equiv\frac{A^{2}}{\omega}\nonumber
\end{equation}
is the noise spectrum function, with $A$ being the spectrum strength. Here, we have also introduced both a low $\omega_{\rm min}$ and high $\omega_{\rm max}$ frequency cut-offs~\cite{schriefl2006decoherence}. In addition, the bosonic occupation number in thermal equilibrium is reduced as $n(\omega)=1/({\rm exp}(\hbar\omega/k_{B}T)-1)\approx\,k_{B}T/\hbar\omega$ for all the low frequency charge noise mode~\cite{Li_2019}. Following our previous study, we choose $A=20$ MHz~\cite{lirui2018a}, $\omega_{\rm min}=0.01$ Hz~\cite{yoneda2018} and $\omega_{\rm max}=5\times10^5$ Hz~\cite{yoneda2018} in our calculations.

The phase coherence time $T^{*}_{2}$ of the spin qubit is given by $\langle\Gamma(T^{*}_{2})\rangle_{\Theta}=1$, i.e., at time $T^{*}_{2}$, the coherence is reduced from $1$ to $e^{-1}$. We remark that in the short time limit $t<1/\omega_{\rm max}$, the decay factor shown in Eq.~(\ref{Eq_decay}) has a simple expression $\langle\Gamma(t)\rangle_{\Theta}\approx\,A^{2}t^{2}\frac{(x_{\Uparrow}-x_{\Downarrow})^{2}}{2(b_{2}-a)^{2}}{\rm ln}\frac{\omega_{\rm max}}{\omega_{\rm min}}$. Under this circumstance, the phase coherence time can be written as 
\begin{equation}
T^{*}_{2}\approx\sqrt{\frac{2(b_{2}-a)^{2}}{(x_{\Uparrow}-x_{\Downarrow})^{2}A^{2}{\rm ln}\frac{\omega_{\rm max}}{\omega_{\rm min}}}}.\label{Eq_T2}
\end{equation}

In Fig.~\ref{Fig_dephasing}, we show the coherence time $T^{*}_{2}$ as a function of the asymmetrical potential parameters of the DQD. As expected, as a consequence of the results given in Fig.~\ref{Fig_Xoperator}, the phase coherence time $T^{*}_{2}$ also has a nonmonotonic dependence on the asymmetrical parameter $V_{1}$ or $b_{1}$ of the DQD. This is also supported by the analytical expression of $T^{*}_{2}$ given in Eq.~(\ref{Eq_T2}). The coherence time $T^{*}_{2}$ gets its minimum at the critical potential parameter $V^{c}_{1}$ [see Fig.~\ref{Fig_dephasing}(a)] or $b^{c}_{1}$ [see Fig.~\ref{Fig_dephasing}(b)]. In addition, when the confining potential is symmetrical, i.e., $V_{1}=0$ meV in Fig.~\ref{Fig_dephasing}(a) or $b_{1}=60$ nm in Fig.~\ref{Fig_dephasing}(b), there is no spin dephasing $T^{*}_{2}=\infty$, as is already illustrated in Fig.~\ref{Fig_Xoperator}. From our calculated results shown in the Figs.~\ref{Fig_Xoperator} and~\ref{Fig_dephasing}, we conclude that a little asymmetry in the DQD confining potential can indeed result in a strong spin dephasing.

\section{\label{sec_V}The Magnetic field dependence of dephasing}
\begin{figure}
\includegraphics{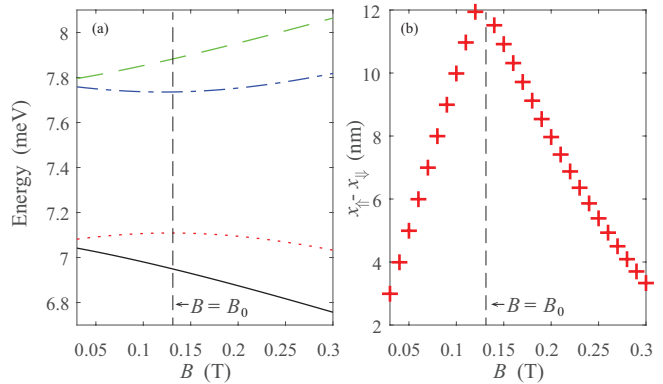}
\caption{\label{Fig_MFdependence}(a) The lowest four energy levels in the DQD as a function of the applied magnetic field $B$ near the anti-crossing point. (b)The difference $x_{\Uparrow}-x_{\Downarrow}$ as a function of the applied magnetic field $B$. Here we have chosen the parameters $V_{1}=0.2$ meV and $b_{1}=60$ nm for both (a) and (b).}	
\end{figure}

\begin{figure}
\includegraphics{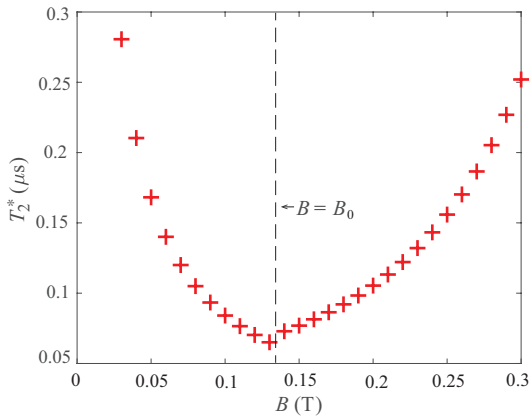}
\caption{\label{Fig_DephasingB}The phase coherence time $T^{*}_{2}$ as a function of the applied magnetic field $B$ near the anti-crossing point of the DQD. Here $V_{1}=0.2$ meV and $b_{1}=60$ nm.}
\end{figure}

It is well-known that there is an anti-crossing structure~\cite{PhysRevB.74.155433} in the energy versus magnetic field plot of a DQD with SOC. Near the anti-crossing point $B_{0}$, the spin degree of freedom of the electron in the DQD is highly hybridized with its orbital degree of freedom, such that a strong electric-dipole spin resonance~\cite{Liu:2018aa} or a strong spin-cavity interaction~\cite{Petersson:2012aa,hu2012strong} is achievable. One interesting question is how does the magnetic field affect the spin dephasing in the DQD, especially near the anti-crossing point.

In many cases of spin pure dephasing, the magnetic field only affects the Zeeman splitting of the electron~\cite{witzel2006quantum,yao2006theory}, thus does not contribute to the spin dephasing. However, in the spin dephasing mechanism mediated by the interplay between the SOC and the asymmetrical confining potential, the magnetic field obviously plays a more complicated role. The magnetic field here not only gives rise to the Zeeman splitting in the DQD, but also contributes to the spin-charge interaction, especially near the anti-crossing point. 

We show the lowest four energy levels in the DQD as a function of the applied magnetic field $B$ in Fig.~\ref{Fig_MFdependence}(a). The level anti-crossing, induced by the SOC in the DQD, can be clearly seen from the figure. Obviously, without SOC ($\alpha=0$), operator $\sigma^{x}$ in Hamiltonian (\ref{Eq_model}) is a conserved quantity, hence there would be only level crossing instead of anti-crossing. The separation of the second and third levels is minimal at the anti-crossing point $B_{0}$ [see Fig.~\ref{Fig_MFdependence}(a)]. The energy gap at the anti-crossing point is very large (about 0.63 meV) due to both the strong SOC and the large effective dot size in the InSb DQD. It is evident that the magnitude of this gap indeed reflects the strength of the SOC in the nanowire material.

We also calculate the difference $x_{\Uparrow}-x_{\Downarrow}$, a quantity reflecting the longitudinal spin-charge interaction, as a function of the applied magnetic field $B$ near the anti-crossing point [see Fig.~\ref{Fig_MFdependence}(b)]. As expected, there is a sharp peak at the anti-crossing point $B_{0}$. This is reasonable, at the anti-crossing point, the hybridization between the spin and orbital degrees of freedom of the electron becomes maximal, such that the longitudinal spin-charge interaction also achieves its maximum. When we tune the magnetic field away from the anti-crossing point, the difference $x_{\Uparrow}-x_{\Downarrow}$ is getting smaller.

It follows that the phase coherence time $T^{*}_{2}$ also has a similar dependence on the magnetic field $B$ [see Fig.~\ref{Fig_DephasingB}], except that the peak structure is changed to a dip structure. This is easy to understand from the approximated expression of $T^{*}_{2}$ given in Eq.~(\ref{Eq_T2}). Before the anti-crossing point $B_{0}$, with the increase of the field, the coherence time $T^{*}_{2}$ decreases, while after the anti-crossing point, $T^{*}_{2}$ is getting larger instead. There is a shortest coherence time $T^{*}_{2}$ at the anti-crossing magnetic field $B_{0}$, where the spin dephasing is strongest.

\section{\label{sec_VI}Discussion and Summary}
Whether there are other mechanisms leading to the longitudinal spin-charge interaction in semiconductor quantum dot is an interesting question. In our paper, the longitudinal spin-charge interaction is mediated by the interplay between the SOC and the asymmetrical confining potential. Here, we give some evidences supporting that our suggested mechanism indeed makes sense. First, if the potential in Hamiltonian (\ref{Eq_model}) is symmetrical, i.e., $V(x)=V(-x)$, then due to the following $Z_{2}$ symmetry $(\sigma^{x}{\mathcal P})H(\sigma^{x}{\mathcal P})=H$, where ${\mathcal P}$ is the parity, the difference $x_{\Uparrow}-x_{\Downarrow}$ is zero~\cite{lirui2018a}. Second, one may want to break the $Z_{2}$ symmetry by introducing a general Zeeman field in the Hamiltonian, i.e., 
\begin{equation}
H=\frac{p^{2}}{2m}+\alpha\sigma^{z}p+\Delta_{1}\sigma^{x}+\Delta_{2}\sigma^{z}+\frac{1}{2}m\omega^{2}x^{2},
\end{equation} 
where we have chosen a harmonic confining potential $V(x)=\frac{1}{2}m\omega^{2}x^{2}$ for illustration. We still can prove $x_{\Uparrow}-x_{\Downarrow}=0$ for this case. It is easy to show the following commutation relation
\begin{equation}
[p,H]=pH-Hp=-i\hbar\,m\omega^{2}x.
\end{equation}
It follows directly
\begin{eqnarray} x_{\Uparrow}&=&\frac{i}{\hbar\,m\omega^{2}}\langle1\Uparrow|pH-Hp|1\Uparrow\rangle=0,\nonumber\\
x_{\Downarrow}&=&\frac{i}{\hbar\,m\omega^{2}}\langle1\Downarrow|pH-Hp|1\Downarrow\rangle=0.
\end{eqnarray}
Hence, there is still no longitudinal spin-charge interaction in this case.

In summary, in this paper we have built explicitly the theory of the $1/f$ charge noise induced spin dephasing in a nanowire DQD. The interplay between the SOC and the asymmetrical confining potential mediates a longitudinal spin-charge interaction in the DQD. The spin dephasing is not monotonically dependent on the degree of the asymmetry of the confining potential. The applied external magnetic field not only gives rise to the Zeeman splitting, but also contributes to the spin-charge interaction, hence the spin is severely dephased near the anti-crossing point $B_{0}$ in the DQD.

\section*{Acknowledgements}
This work is supported by the National Natural Science Foundation of China Grant No.~11404020, the Postdoctoral Science Foundation of China Grant No.~2014M560039, the Project from the Department of Education of Hebei Province Grant No. QN2019057, and the Starting up Foundation from Yanshan University Grant No. BL18043.

\appendix
\begin{widetext}
\section{\label{appendix_a}The derivation of the transcendental equation}
We first solve the continuous spectrum for the bulk Hamiltonian $H_{0}=\frac{p^{2}}{2m}+\alpha\sigma^{z}p+\Delta\sigma^{x}$ in the absence of the quantum dot confining potential. The explicit dispersion relation and the bulk wavefunctions can be found elsewhere~\cite{lirui2018the,lirui2018a}. Now, we can write the eigenfunction in the quantum dot as follows
\begin{equation}
\Psi(x)=\left\{\begin{array}{cc}
c_{1}\left(\begin{array}{c}
\cos\frac{\theta_{1}}{2}\\
\sin\frac{\theta_{1}}{2}
\end{array}\right)e^{ik_{1}x}+c_{2}\left(\begin{array}{c}
\sin\frac{\theta_{1}}{2}\\
\cos\frac{\theta_{1}}{2}
\end{array}\right)e^{-ik_{1}x}+c_{3}\left(\begin{array}{c}
\sin\frac{\theta_{2}}{2}\\
-\cos\frac{\theta_{2}}{2}
\end{array}\right)e^{ik_{2}x}+c_{4}\left(\begin{array}{c}
\cos\frac{\theta_{2}}{2}\\
-\sin\frac{\theta_{2}}{2}
\end{array}\right)e^{-ik_{2}x},&{\rm I}\\
c_{5}\left(\begin{array}{c}
1\\
Re^{i\Phi}
\end{array}\right)e^{ik_{x}x-k_{y}x}+c_{6}\left(\begin{array}{c}
Re^{-i\Phi}\\
1
\end{array}\right)e^{-ik_{x}x-k_{y}x}+c_{7}\left(\begin{array}{c}
1\\
Re^{-i\Phi}
\end{array}\right)e^{ik_{x}x+k_{y}x}+c_{8}\left(\begin{array}{c}
Re^{i\Phi}\\
1
\end{array}\right)e^{-ik_{x}x+k_{y}x},&{\rm II}\\
c_{9}\left(\begin{array}{c}
\cos\frac{\theta_{3}}{2}\\
\sin\frac{\theta_{3}}{2}
\end{array}\right)e^{ik_{3}x}+c_{10}\left(\begin{array}{c}
\sin\frac{\theta_{3}}{2}\\
\cos\frac{\theta_{3}}{2}
\end{array}\right)e^{-ik_{3}x}+c_{11}\left(\begin{array}{c}
\sin\frac{\theta_{4}}{2}\\
-\cos\frac{\theta_{4}}{2}
\end{array}\right)e^{ik_{4}x}+c_{12}\left(\begin{array}{c}
\cos\frac{\theta_{4}}{2}\\
-\sin\frac{\theta_{4}}{2}
\end{array}\right)e^{-ik_{4}x},&{\rm III}
\end{array}\right.\label{Eq_eigenfunction}
\end{equation}
where
\begin{eqnarray}
k_{1,2}&=&\sqrt{2}m\alpha\sqrt{1+\frac{E}{m\alpha^{2}}\mp\sqrt{1+2\frac{E}{m\alpha^{2}}+\frac{\Delta^{2}}{m^{2}\alpha^{2}}}},~\theta_{1,2}=\arctan\left(\Delta/(\alpha\,k_{1,2})\right),\nonumber\\
k_{x,y}&=&m\alpha\sqrt{\pm1\pm\frac{E-V_{0}}{m\alpha^{2}}+\sqrt{\frac{(E-V_{0})^{2}-\Delta^{2}}{m^{2}\alpha^{4}}}},~R\cos\Phi=-\frac{m\alpha^{2}+\alpha\,k_{x}}{\Delta},~R\sin\Phi=-\frac{k_{x}k_{y}+m\alpha\,k_{y}}{m\Delta},\nonumber\\
k_{3,4}&=&\sqrt{2}m\alpha\sqrt{1+\frac{E-V_{1}}{m\alpha^{2}}\mp\sqrt{1+2\frac{E-V_{1}}{m\alpha^{2}}+\frac{\Delta^{2}}{m^{2}\alpha^{2}}}},~\theta_{3,4}=\arctan\left(\Delta/(\alpha\,k_{3,4})\right).
\end{eqnarray}
We totally have 12 coefficients $c_{i}$ ($i=1,\ldots,12$) to be determined. The boundary condition (\ref{Eq_boundary}) actually has 12 subequations. Therefore, when $\Psi(x)$ in Eq.~(\ref{Eq_boundary}) is replaced with the above expanded form, we obtain the following matrix equation 
\begin{equation}
{\bf M}\cdot{\bf C}=0.\label{eq_matrixarray}
\end{equation}
where ${\bf M}$ is a $12\times12$ matrix and ${\bf C}=[c_{1},~c_{2},~\ldots~c_{12}]^{\rm T}$. If the above equation array has non-trivial solutions, the determinant of matrix ${\bf M}$ must be zero, i.e.,
\begin{equation}
{\rm det}({\bf M})=0.
\end{equation} 
This transcendental equation is actually an implicit equation of the eigen-energy $E$. Solving this equation, we obtain the energy spectrum of the DQD. Once an eigen-value $E_{n}$ is obtained, we can solve the coefficients $c^{n}_{i}$ (i=1,\ldots,12) via Eq.~(\ref{eq_matrixarray}). Hence, the corresponding eigenfunctions $\Psi_{n}(x)$ with eigen-value $E_{n}$ can be determined from Eq.~(\ref{Eq_eigenfunction}).

The detailed matrix elements of ${\bf M}$ read
\begin{eqnarray}
M_{1,1}&=&e^{-ik_{1}b_{2}}\cos\frac{\theta_{1}}{2},~M_{1,2}=e^{ik_{1}b_{2}}\sin\frac{\theta_{1}}{2},~M_{1,3}=e^{-ik_{2}b_{2}}\sin\frac{\theta_{2}}{2},~M_{1,4}=e^{ik_{2}b_{2}}\cos\frac{\theta_{2}}{2},\nonumber\\
M_{1,5}&=&M_{1,6}=M_{1,7}=M_{1,8}=M_{1,9}=M_{1,10}=M_{1,11}=M_{1,12}=0,\nonumber\\
M_{2,1}&=&e^{-ik_{1}b_{2}}\sin\frac{\theta_{1}}{2},~M_{2,2}=e^{ik_{1}b_{2}}\cos\frac{\theta_{1}}{2},~M_{2,3}=-e^{-ik_{2}b_{2}}\cos\frac{\theta_{2}}{2},~M_{2,4}=-e^{ik_{2}b_{2}}\sin\frac{\theta_{2}}{2},\nonumber\\
M_{2,5}&=&M_{2,6}=M_{2,7}=M_{2,8}=M_{2,9}=M_{2,10}=M_{2,11}=M_{2,12}=0,\nonumber\\
M_{3,1}&=&e^{-ik_{1}a}\cos\frac{\theta_{1}}{2},~M_{3,2}=e^{ik_{1}a}\sin\frac{\theta_{1}}{2},~M_{3,3}=e^{-ik_{2}a}\sin\frac{\theta_{2}}{2},~M_{3,4}=e^{ik_{2}a}\cos\frac{\theta_{2}}{2},~M_{3,5}=-e^{-ik_{x}a+k_{y}a},\nonumber\\
M_{3,6}&=&-Re^{-i\Phi+ik_{x}a+k_{y}a},~M_{3,7}=-e^{-ik_{x}a-k_{y}a},~M_{3,8}=-Re^{i\Phi+ik_{x}a-k_{y}a},~M_{3,9}=M_{3,10}=M_{3,11}=M_{3,12}=0,\nonumber\\ 
M_{4,1}&=&e^{-ik_{1}a}\sin\frac{\theta_{1}}{2},~M_{4,2}=e^{ik_{1}a}\cos\frac{\theta_{1}}{2},~M_{4,3}=-e^{-ik_{2}a}\cos\frac{\theta_{2}}{2},~M_{4,4}=-e^{ik_{2}a}\sin\frac{\theta_{2}}{2},\nonumber\\
M_{4,5}&=&-Re^{i\Phi-ik_{x}a+k_{y}a},~M_{4,6}=-e^{ik_{x}a+k_{y}a},~M_{4,7}=-Re^{-i\Phi-ik_{x}a-k_{y}a},~M_{4,8}=-e^{ik_{x}a-k_{y}a},\nonumber\\
M_{4,9}&=&M_{4,10}=M_{4,11}=M_{4,12}=0,\nonumber\\
M_{5,1}&=&M_{5,2}=M_{5,3}=M_{5,4}=0,~M_{5,5}=e^{ik_{x}a-k_{y}a},~M_{5,6}=Re^{-i\Phi-ik_{x}a-k_{y}a},~M_{5,7}=e^{ik_{x}a+k_{y}a},\nonumber\\
M_{5,8}&=&Re^{i\Phi-ik_{x}a+k_{y}a},~M_{5,9}=-e^{ik_{3}a}\cos\frac{\theta_{3}}{2},~M_{5,10}=-e^{-ik_{3}a}\sin\frac{\theta_{3}}{2},\nonumber\\
M_{5,11}&=&-e^{ik_{4}a}\sin\frac{\theta_{4}}{2},~M_{5,12}=-e^{-ik_{4}a}\cos\frac{\theta_{4}}{2},\nonumber\\
M_{6,1}&=&M_{6,2}=M_{6,3}=M_{6,4}=0,~M_{6,5}=Re^{i\Phi+ik_{x}a-k_{y}a},~M_{6,6}=e^{-ik_{x}a-k_{y}a},~M_{6,7}=Re^{-i\Phi+ik_{x}a+k_{y}a},\nonumber\\
M_{6,8}&=&e^{-ik_{x}a+k_{y}a},~M_{6,9}=-e^{ik_{3}a}\sin\frac{\theta_{3}}{2},~M_{6,10}=-e^{-ik_{3}a}\cos\frac{\theta_{3}}{2},~M_{6,11}=e^{ik_{4}a}\cos\frac{\theta_{4}}{2},~M_{6,12}=e^{-ik_{4}a}\sin\frac{\theta_{4}}{2},\nonumber\\
M_{7,1}&=&M_{7,2}=M_{7,3}=M_{7,4}=M_{7,5}=M_{7,6}=M_{7,7}=M_{7,8}=0,~M_{7,9}=e^{ik_{3}b_{1}}\cos\frac{\theta_{3}}{2},~M_{7,10}=e^{-ik_{3}b_{1}}\sin\frac{\theta_{3}}{2},\nonumber\\
M_{7,11}&=&e^{ik_{4}b_{1}}\sin\frac{\theta_{4}}{2},~M_{7,12}=e^{-ik_{4}b_{1}}\cos\frac{\theta_{4}}{2},\nonumber\\
M_{8,1}&=&M_{8,2}=M_{8,3}=M_{8,4}=M_{8,5}=M_{8,6}=M_{8,7}=M_{8,8}=0,~M_{8,9}=e^{ik_{3}b_{1}}\sin\frac{\theta_{3}}{2},~M_{8,10}=e^{-ik_{3}b_{1}}\cos\frac{\theta_{3}}{2},\nonumber\\
M_{8,11}&=&-e^{ik_{4}b_{1}}\cos\frac{\theta_{4}}{2},~M_{8,12}=-e^{-ik_{4}b_{1}}\sin\frac{\theta_{4}}{2},\nonumber\\
M_{9,1}&=&ik_{1}e^{-ik_{1}a}\cos\frac{\theta_{1}}{2},~M_{9,2}=-ik_{1}e^{ik_{1}a}\sin\frac{\theta_{1}}{2},~M_{9,3}=ik_{2}e^{-ik_{2}a}\sin\frac{\theta_{2}}{2},~M_{9,4}=-ik_{2}e^{ik_{2}a}\cos\frac{\theta_{2}}{2},\nonumber\\
M_{9,5}&=&-(ik_{x}-k_{y})e^{-ik_{x}a+k_{y}a},~M_{9,6}=(ik_{x}+k_{y})Re^{-i\Phi+ik_{x}a+k_{y}a},~M_{9,7}=-(ik_{x}+k_{y})e^{-ik_{x}a-k_{y}a},\nonumber\\
M_{9,8}&=&(ik_{x}-k_{y})Re^{i\Phi+ik_{x}a-k_{y}a},~M_{9,9}=M_{9,10}=M_{9,11}=M_{9,12}=0,\nonumber\\
M_{10,1}&=&ik_{1}e^{-ik_{1}a}\sin\frac{\theta_{1}}{2},~M_{10,2}=-ik_{1}e^{ik_{1}a}\cos\frac{\theta_{1}}{2},~M_{10,3}=-ik_{2}e^{-ik_{2}a}\cos\frac{\theta_{2}}{2},~M_{10,4}=ik_{2}e^{ik_{2}a}\sin\frac{\theta_{2}}{2},\nonumber\\
M_{10,5}&=&-(ik_{x}-k_{y})Re^{i\Phi-ik_{x}a+k_{y}a},~M_{10,6}=(ik_{x}+k_{y})e^{ik_{x}a+k_{y}a},~M_{10,7}=-(ik_{x}+k_{y})Re^{-i\Phi-ik_{x}a-k_{y}a},\nonumber\\
M_{10,8}&=&(ik_{x}-k_{y})e^{ik_{x}a-k_{y}a},~M_{10,9}=M_{10,10}=M_{10,11}=M_{10,12}=0,\nonumber\\
M_{11,1}&=&M_{11,2}=M_{11,3}=M_{11,4}=0,~M_{11,5}=(ik_{x}-k_{y})e^{ik_{x}a-k_{y}a},~M_{11,6}=(-ik_{x}-k_{y})Re^{-i\Phi-ik_{x}a-k_{y}a},\nonumber\\
M_{11,7}&=&(ik_{x}+k_{y})e^{ik_{x}a+k_{y}a},~M_{11,8}=(-ik_{x}+k_{y})Re^{i\Phi-ik_{x}a+k_{y}a},~M_{11,9}=-ik_{3}e^{ik_{3}a}\cos\frac{\theta_{3}}{2},\nonumber\\
M_{11,10}&=&ik_{3}e^{-ik_{3}a}\sin\frac{\theta_{3}}{2},~M_{11,11}=-ik_{4}e^{ik_{4}a}\sin\frac{\theta_{4}}{2},~M_{11,12}=ik_{4}e^{-ik_{4}a}\cos\frac{\theta_{4}}{2},\nonumber\\
M_{12,1}&=&M_{12,2}=M_{12,3}=M_{12,4}=0,~M_{12,5}=(ik_{x}-k_{y})Re^{i\Phi+ik_{x}a-k_{y}a},~M_{12,6}=(-ik_{x}-k_{y})e^{-ik_{x}a-k_{y}a},\nonumber\\
M_{12,7}&=&(ik_{x}+k_{y})Re^{-i\Phi+ik_{x}a+k_{y}a},~M_{12,8}=(-ik_{x}+k_{y})e^{-ik_{x}a+k_{y}a},~M_{12,9}=-ik_{3}e^{ik_{3}a}\sin\frac{\theta_{3}}{2},\nonumber\\
M_{12,10}&=&ik_{3}e^{-ik_{3}a}\cos\frac{\theta_{3}}{2},~M_{12,11}=ik_{4}e^{ik_{4}a}\cos\frac{\theta_{4}}{2},~M_{12,12}=-ik_{4}e^{-ik_{4}a}\sin\frac{\theta_{4}}{2}.
\end{eqnarray}
\end{widetext}

\bibliographystyle{iopart-num}
\bibliography{RuiLiRef}
\end{document}